%Paper: hep-th/9405050
%From: juriev@physique.ens.fr (JURIEV Denis)
%Date: Mon, 9 May 94 11:42:04 +0200
%Date (revised): Tue, 10 May 94 09:06:38 +0200
%Date (revised): Mon, 16 May 94 12:04:25 +0200
%Date (revised): Wed, 21 Sep 94 10:22:35 +0200
%Date (revised): Tue, 27 Sep 94 08:17:41 +0100
%Date (revised): Sat, 1 Oct 94 14:17:56 +0100

\input amstex
\documentstyle{amsppt}
\magnification=1200
\leftheadtext{D.Juriev}
\rightheadtext{TOPICS IN HIDDEN SYMMETRIES}
\NoBlackBoxes
%%%%%%%%%%%%%%%%%%%%%%%%%%%%%%%%%%%%%%%%%%%%%%%%%%%%%%%%%%%%%%%%%%%%
 \define\FA{\frak A} \define\CA{\Cal A}

  \define\CF{\Cal F}
\define\Fg{\frak g}

\define\Fl{\frak l}  
  \define\CM{\Cal M}
  
\define\Fo{\frak o}

\define\Fs{\frak s}  \define\CS{\Cal S}
  
  \define\CU{\Cal U}

%-----

\define\AC{\Bbb C}

\define\AQ{\Bbb Q}   
\define\AR{\Bbb R}

\define\AZ{\Bbb Z}   
%-----

\define\TO{\text{\tt O}}

%-----
 \define\vtheta\vartheta

\define\gl{\lambda}

%%%%%%%%%%%%%%%%%%%%%%%%%%%%%%%%%%%%%%%%%%%%%%%%%%%%%%%%%%%%%%%%%%%%%%%%%%

\define\End{\operatorname{End}}

\define\Der{\operatorname{Der}}

\define\ad{\operatorname{ad}}

%-------------------------------------------------------------------------
\define\pd#1#2{\partial_{#1}^{#2}}
\define\p#1{\partial_{#1}}
\define\fpd#1#2{(\frac{\partial}{\partial {#1}})^{#2}}
\define\fp#1{\frac{\partial}{\partial {#1}}}
%-------------------------------------------------------------------------
\define\Gr#1#2#3{#1(#2,#3)}
\define\GrC#1#2{\Gr#1#2{\AC}}
\define\GrR#1#2{\Gr#1#2{\AR}}
\define\GrQ#1#2{\Gr#1#2{\AQ}}
\define\GrZ#1#2{\Gr#1#2{\AZ}}
\define\vGr#1#2#3{#1_{#2}(#3)}
\define\vGrC#1#2{\vGr#1#2{\AC}}
\define\vGrR#1#2{\vGr#1#2{\AR}}
\define\vGrQ#1#2{\vGr#1#2{\AQ}}
\define\vGrZ#1#2{\vGr#1#2{\AZ}}
%-----

%-----
 
\define\sla{\operatorname{\Fs\Fl}}

\define\soa{\operatorname{\Fs\Fo}} 
%%%%%%%%%%%%%%%%%%%%%%%%%%%%%%%%%%%%%%%%%%%%%%%%%%%%%%%%%%%%%%%%%%%%%%
\define\sLtwo{\GrC{\sla}2}

\define\sothree{\GrC{\soa}3}
\define\RW{\operatorname{\Cal R\Cal W}}
\define\cRW{\widehat{\RW}}
\define\RWtwo{\RW(\sLtwo)}

\define\RWtwop#1{\RW(\sLtwo;#1)}
\define\cRWtwop#1{\cRW(\sLtwo;#1)}
\define\pz{\p z}
\define\pzs{\pd z 2}
\define\pzt{\pd z 3}
\define\Vect{\operatorname{Vect}}
\define\DOP{\operatorname{DOP}}
\document
\qquad\qquad\qquad\qquad\qquad\qquad\qquad\qquad\qquad\qquad\qquad\qquad
\qquad\qquad hep-th/9405050

\

\

\topmatter
\title TOPICS IN HIDDEN SYMMETRIES
\endtitle
\author\eightpoint Denis Juriev\footnote{\ On leave from Mathematical Division,
Research Institute for System Studies, Russian Academy of Sciences, Moscow,
Russia (e-mail: juriev\@systud.msk.su)\newline}
\centerline{}
\centerline{}
\centerline{}
\centerline{Laboratoire de Physique Th\'eorique de l'\'Ecole Normale
Sup\'erieure,}
24 rue Lhomond, 75231 Paris Cedex 05, France\footnote{\ Unit\'e Propre du
Centre
National de la Recherche Scientifique associ\'ee \`a l'\'Ecole Normale
Sup\'erieure et \`a l'Universit\'e de Paris--Sud\newline}
\centerline{E mail: juriev\@physique.ens.fr}
\endauthor
\abstract These three topics are an attempt to explicate some curiosities of
{\sl the inverse problem of representation theory} (i.e. having a set of
operators to describe the "correct" algebraic object, which is represented by
them) on simple examples related to the Lie algebra $\sLtwo$.
\endabstract
\endtopmatter

\head INTRODUCTION \endhead

The representation theory has, at least, two faces. The first one is related
to \sl the direct problem of the representation theory \rm i.e. having an
abstract object to describe its representations. Being in general completed
for classical structures (Lie groups and Lie (super)algebras) this face has
now its renaissance provoked by the discovering of a large scope of new
algebraic structures in the modern quantum field theory (quantum groups,
Zamolodchikov algebras, operator algebras of quantum field theory and their
variations, W--algebras and their generalizations, bordism categories and
trains of finite and infinite dimensional groups, homotopy Lie algebras,
Batalin--Vilkovisky algebras, etc.). The second face is related to \sl the
inverse problem of the representation theory\rm: having a set of operators to
describe the "correct" algebraic object, which is represented by them. It
should be marked that the most of new algebraic structures mentioned above was
discovered as solutions of such inverse problem. The second face being very
popular among physicists has no attracted a lot of attention of
mathematicians. These topics on hidden symmetries are an attempt to explicate
some curiosities of this face on simple examples related to the Lie algebra
$\sLtwo$.

The author is glad to thank Prof.Dr.A.N.Rudakov and Mathematical Division,
Research Institute for System Studies for a beautiful scientific atmosphere
as well as  Prof.Dr.J.-L.Gervais and Laboratoire de Physique Th\'eorique de
l'\'Ecole
Normale Sup\'erieure for very perfect conditions.

\head TOPIC ONE: SETTING HIDDEN SYMMETRIES FREE BY THE NONCOMMUTATIVE
SEMICUBIC MAPPING\endhead

As it was marked above one of the main problems to deal with hidden symmetries
is to set them free "correctly", i.e. to find a "correct" algebraic structure,
which is represented by hidden symmetries. It is very convenient to consider
this problem in the framework of noncommutative geometry [1]. Such approach to
hidden symmetries in Verma modules over $\sLtwo$ was adopted in [2], where
hidden symmetries were set free by the noncommutative Veronese mapping.
Another way to set hidden symmetries free, related to the noncommutative
semicubic mapping, is described in the present topic.

\definition{Definition 1}

{\bf A} [2]. Let $\Fg$ be a Lie algebra and $\CA$ be an associative algebra
such
that $\Fg\subset\Der(\CA)$; a linear subspace $V$ of $\CA$ is called {\it a
space of hidden symmetries\/} iff (1) $V$ is a $\Fg$--submodule of $\CA$, (2)
the Weyl symmetrization defines a surjection $W:S^{\cdot}(V)\mapsto\CA$ (the
elements of $V$ are called {\it hidden symmetries with respect to $\Fg$}). An
associative algebra $\CF$ such that $\Fg\subset\Der(\CF)$ is called {\it an
algebra of the set free hidden symmetries\/} iff (1) $\CF$ is generated by
$V$, (2) there exists a $\Fg$--equivariant epimorphism of algebras
$\CF\mapsto\CA$, (3) the Weyl symmetrization defines an isomorphism
$S^{\cdot}(V)\mapsto\CF$ of $\Fg$--modules.

Let $\Fg$ be a Lie algebra, $V$ be a certain $\Fg$--module, $\CA_s$
be a family of associative algebras, parametrized by $s\in\CS$ such that
$\Fg\subset\Der(\CA_s)$, $\pi_s:V\mapsto\CA_s$ be a family of
$\Fg$--equivariant imbeddings such that $\pi_s(V)$ is a space of hidden
symmetries in $\CA_s$ with respect to $\Fg$ for a generic $s$ from $\CS$.  An
associative algebra $\CF$ is called {\it an algebra of the
$\CA_{s,s\in\CS}$--universally set free hidden symmetries\/} iff $\CF$ is an
algebra of the set free hidden symmetries corresponding to $V\simeq\pi_s(V)$
for generic $\CA_s$ ($s\in\CS$).

{\bf B.} Let $V$ be a space of hidden symmetries in algebra $\CA$ with respect
to the Lie algebra $\Fg$; hidden symmetries from $V$ will be called of {\it
semicubic type\/} iff there exist (1) the nontrivial decomposition of $V$ into
the direct sum $V_2\oplus V_3$ of its subspaces $V_2$ and $V_3$, (2) an
irreducible $\Fg$--module $V_1$ such that the epimorphisms $S^i(V_1)\mapsto
V_i$ ($i=2,3$) of $\Fg$--modules are defined; in this case the mapping
$\bigoplus_{i=2,3}S^i(V_1)\mapsto\CF$, a composition of mappings
$S^i(V_1)\mapsto V_i$ and the imbedding of $V$ into $\CF$, is called {\it the
noncommutative semicubic mapping}.
\enddefinition

\proclaim{Theorem 1}
The tensor operators of type $\pi_2$ and $\pi_3$ in the Verma module $V_h$
over the Lie algebra $\sLtwo$ ($\pi_i$ is a finite--dimensional representation
of $\sLtwo$ of dimension $2i+1$) form a space of hidden symmetries of
semicubic type; there exist exactly one quadratic (homogeneous)
algebra\footnote{\ Such algebras form a very natural class of associative
algebras. Their exploration was initiated independently and simultaneously by
several authors, the ref [3] should be marked as less known than others. One
should also find ref [4] among more recent ones.} $\FA_h$ of the set free
hidden
symmetries for each $h$.
\endproclaim

Let $L_i$ be a basis in $\sLtwo$ such that $$[L_i,L_j]=(i-j)L_{i+j},
\quad (i,j=-1,2,3)$$ and $d^k_j$ ($-k\le j\le k$) be basises in $\pi_k$,
in which the $\sLtwo$--action has the form $$L_i(d^k_j)=(ki-j)d^k_{i+j}.$$ The
corresponding tensor operators in the Verma modules $V_h$ will be denoted by
the capitals. If the Verma module $V_h$ is realized in the space $\AC[z]$ of
polynomials of a complex variable $z$, where the generators of $\sLtwo$ act as
$$L_{-1}=z,\quad L_0=z\pz+h,\quad L_1=z\pzs+2h\pz,$$ then
the tensor operators $D^k_i$ ($k=2,3$) are defined by the formulas
$$\align
D_{-2}^2&=z^2\\
D_{-1}^2&=z(\xi+h+\tfrac12)\\
D_0^2&=\xi^2+2h\xi+\tfrac13h(2h+1)\\
D_1^2&=(\xi+2h)(\xi+h+\tfrac12)\pz\\
D_2^2&=(\xi+2h)(\xi+2h+1)\pzs\\
     & \\ \allowdisplaybreak
D_{-3}^3&=z^3\\
D_{-2}^3&=z^2(\xi+h+1)\\
D_{-1}^3&=z(\xi^2+(2h+1)\xi+\tfrac25(h+1)(2h+1))\\
D_0^3&=\xi^3+3h\xi^2+(2h^2+\tfrac15(h+1)(2h+1))\xi+\tfrac15 h(h+1)(2h+1)\\
D_1^3&=(\xi+2h)(\xi^2+(2h+1)\xi+\tfrac25(h+1)(2h+1))\pz\\
D_2^3&=(\xi+2h)(\xi+2h+1)(\xi+h+1)\pzs\\
D_3^3&=(\xi+2h)(\xi+2h+1)(\xi+2h+2)\pzt
\endalign
$$
where $\xi=z\pz$.

\demo{Proof of the Theorem} Let's denote the mapping $d^k_i\mapsto D^k_i$ by
$\TO$ then the tensor product
$$\pi_i\otimes\pi_j=\bigoplus_{k=|i-j|}^{i+j}\pi_k$$
maybe divided in two parts:
$$\pi_i\otimes\pi_j=\pi_i\vee\pi_j\oplus\pi_i\wedge\pi_j,$$ where
$$\TO(\pi_i\vee\pi_j)=[\TO(\pi_i),\TO(\pi_j)]_+,\quad
\TO(\pi_i\wedge\pi_j)=[\TO(\pi_i),\TO(\pi_j)]_-$$ or explicitely $$\align
\pi_i\vee\pi_j&=\bigoplus_{k=|i-j|, i+j-k\in 2\AZ}^{i+j}\pi_k,\\
\pi_i\wedge\pi_j&=\bigoplus_{k=|i-j|, i+j-k\in 2\AZ+1}^{i+j}\pi_k.
\endalign$$ The $\sLtwo$--invariant structure of the homogeneous quadratic
algebra is defined by the expression of commutators $[\pi_2,\pi_2]_-$,
$[\pi_2,\pi_3]_-$ and $[\pi_3,\pi_3]_-$ via anticommutators $[\pi_2,\pi_3]_+$,
$[\pi_2,\pi_2]_++[\pi_3,\pi_3]_+$ and $[\pi_2,\pi_3]_+$, respectively (here
$\pi_i$ is a collective notation for the set free hidden symmetries, which
transforms accordingly to such representation). Such expression is
corresponded to the imbeddings
$$\align
\pi_2\wedge\pi_2&=\pi_3\oplus\pi_1\subseteq\pi_5\oplus\pi_3\oplus\pi_1=
\pi_2\vee\pi_3,\\
\pi_2\wedge\pi_3
&=\pi_4\oplus\pi_2\subseteq(\pi_4\oplus\pi_2)+(\pi_6\oplus\pi_4\oplus\pi_2)
=\pi_2\vee\pi_2+\pi_3\vee\pi_3,\\
\pi_3\wedge\pi_3&=\pi_5\oplus\pi_3\oplus\pi_1=\pi_2\vee\pi_3;
\endalign
$$
so the general commutation relations contain 4 indeterminate constants, two of
them (the first and the fourth) maybe normalized by the scaling redefinition
of tensor operators $D^k_i$: $D^2_i\to\gl_2 D^2_i$, $D^3_i\to\gl_3 D^3_i$,
the second and the third constants are fixed by the Jacobi identities for
triples from $\pi_2\otimes\pi_2\otimes\pi_3$ and
$\pi_3\otimes\pi_3\otimes\pi_2$; Jacobi identities for triples from
$\pi_2\otimes\pi_2\otimes\pi_2$ and $\pi_3\otimes\pi_3\otimes\pi_3$ do not put
any new conditions
\qed\enddemo

\remark{Remark} Algebras $\FA_h$ are not isomorphic to each other for
different $h$ and there is no any quadratic (homogeneous as well as
non--homogeneous) algebra, which set the described hidden symmetries
$\End(V_h)$--universally.
\endremark

\

{\bf Hypothesis 1.} The statement of the Theorem 1 remains true after the
change
of $\pi_2$ and $\pi_3$ on $\pi_n$ and $\pi_{n+1}$ ($n\ge 3$).

\

{\bf Question 1:} Are the algebras $\FA_h$ of the Theorem 1 Koszul algebras
[5]?

\remark{Remark} It seems that the investigation of the set free hidden
symmetries, defined by tensor operators, may enlight some additional relations
in the theory of Clebsch--Gordan coefficients [6].
\endremark

\head TOPIC TWO: MHO--ALGEBRAS\endhead

\definition{Definition 2} Let $\Fg$ be a Lie algebra and $W$ be its
(irreducible) representation. {\it Mho--algebra\/} $\mho(\Fg,W)$ is an
associative algebra such that (1) $\CU(\Fg)$ is a subalgebra of $\mho(\Fg,W)$,
so $\Fg$ naturally acts in $\mho(\Fg,W)$ by the adjoint action, (2) there
exist a $\Fg$--equivariant imbedding of $W$ into $\mho(\Fg,W)$, so one may
consider $W$ as a subspace of $\mho(\Fg,W)$, (3) the $\Fg$--equivariant
imbedding of $W$ into $\mho(\Fg,W)$ is extented to a $\Fg$--equivariant
imbedding of $S^{\cdot}(W)$ into $\mho(\Fg,W)$ defined by the Weyl
symmetrization mapping, so one may consider $S^{\cdot}(W)$ as a subspace of
$\mho(\Fg,W)$; (4) the isomorphisms of $\Fg$--modules $\mho(\Fg,W)$ and
$S^{\cdot}(\Fg)\otimes S^{\cdot}(W)$ holds, here the isomorphism of subalgebra
$\CU(\Fg)$ of $\mho(\Fg,W)$ and $S^{\cdot}(\Fg)$ as $\Fg$--modules is used;
(5) let's fix an arbitrary basis $w_l$ in $W$ then the commutator of two
elements of the basis in the algebra $\mho(\Fg,W)$ maybe represented as
$[w_i,w_j]=f^k_{ij}w_k$, where the "noncommutative structural functions"
$f^k_{ij}$ are certain
elements of $\CU(\Fg)$.
\enddefinition

The notation $\mho(\Fg,W)$ should symbolize an analogy between mho--algebras
and universal envelopping algebras.

\proclaim{Theorem 2} There exists exactly one mho--algebra
$\mho(\sLtwo,\pi_n)$ ($n=1, 2, 3$) with "noncommutative structural functions"
$f^k_{ij}$ of degree $n-1$, which is an algebra of $\End(V_h)$--universally set
free hidden symmetries.
\endproclaim

In the case $n=1$ the mho--algebra $\mho(\sLtwo,\pi_n)$ is the squashed
tensor product of two copies of $\CU(\sLtwo)$ (i.e. the universal envelopping
algebra of the squashed sum of two copies of $\sLtwo$, the first copy acts on
the second one by the adjoint action). In the case $n=2$ the mho--algebra
$\mho(\sLtwo,\pi_n)$ is just the Racah--Wigner algebra $\RWtwo$ of [2]. Let's
describe the mho--algebra $\mho(\sLtwo,\pi_3)$ explicitely.

Let's fix a basis $u_k$ ($-3\le k\le 3$) in $\pi_3$ such that
$L_i(u_k)=(3i-k)u_{i+k}$. Let's also introduce the elements $\Omega_j$
($-2\le j\le 2$; $[L_i,\Omega_j]=(2i-j)\Omega_{i+j}$) in $\CU(\sLtwo)$:
$$\align\Omega_{-2}=L_{-1}^2,\quad
\Omega_{-1}=&\tfrac12(L_0L_{-1}+L_{-1}L_0),\quad
\Omega_0=\tfrac16(L_1L_{-1}+4L^2_0+L_{-1}L_1),\\
&\Omega_1=\tfrac12(L_1L_0+L_0L_1),\quad \Omega_2=L^2_1.\endalign$$
Let's also introduce the expressions $A_j$ ($-5\le j\le 5$;
$[L_i,A_j]=(5i-j)A_{i+j}$), $B_j$ ($-3\le j\le 3$; $[L_i,B_j]=(3i-j)B_{i+j}$),
$C_j$ ($j=-1,0,1$, $[L_i,C_j]=(i-j)C_{i+j}$):
$$\align
A_{-5}&=-\Omega_{-2}u_{-3}\\
A_{-4}&=-\tfrac15(2\Omega_{-1}u_{-3}+3\Omega_{-2}u_{-2})\\
A_{-3}&=-\tfrac1{15}(2\Omega_0u_{-3}+8\Omega_{-1}u_{-2}+5\Omega_{-2}u_{-1})\\
A_{-2}&=-\tfrac1{30}(\Omega_1u_{-3}+9\Omega_0u_{-2}+15\Omega_{-1}u_{-1}+
5\Omega_{-2}u_0)\\
A_{-1}&=-\tfrac1{210}(\Omega_2u_{-3}+24\Omega_1u_{-2}+90\Omega_0u_{-1}+
80\Omega_{-1}u_0+15\Omega_{-2}u_1)\\
A_0&=-\tfrac1{42}(\Omega_2u_{-2}+10\Omega_1u_{-1}+20\Omega_0u_0+
10\Omega_{-1}u_1+\Omega_{-2}u_2)\\
A_1&=-\tfrac1{210}(15\Omega_2u_{-1}+80\Omega_1u_0+90\Omega_0u_1+
24\Omega_{-1}u_2+\Omega_{-2}u_3)\\
A_2&=-\tfrac1{30}(5\Omega_2u_0+15\Omega_1u_1+9\Omega_0u_2+\Omega_{-1}u_3)\\
A_3&=-\tfrac1{15}(5\Omega_2u_1+8\Omega_1u_2+2\Omega_0u_3)\\
A_4&=-\tfrac15(3\Omega_2u_2+2\Omega_1u_3),\\
A_5&=-\Omega_2u_3\\
   &\\ \allowdisplaybreak
B_{-3}&=\Omega_0u_{-3}-2\Omega_{-1}u_{-2}+\Omega_{-2}u_{-1}\\
B_{-2}&=\tfrac13(\Omega_1u_{-3}-3\Omega_{-1}u_{-1}+2\Omega_{-2}u_0)\\
B_{-1}&=\tfrac1{15}(\Omega_2u_{-3}+6\Omega_1u_{-2}-9\Omega_0u_{-1}-
4\Omega_{-1}u_0+6\Omega_{-2}u_1)\\
B_0&=\tfrac15(\Omega_2u_{-2}+\Omega_1u_{-1}-4\Omega_0u_0+\Omega_{-1}u_1+
\Omega_{-2}u_2)\\
B_1&=\tfrac1{15}(6\Omega_2u_{-1}-4\Omega_1u_0-9\Omega_0u_1+6\Omega_{-1}u_2+
\Omega_{-2}u_3)\\
B_2&=\tfrac13(2\Omega_2u_0-3\Omega_1u_1+\Omega_{-1}u_3)\\
B_3&=\Omega_2u_1-2\Omega_1u_2+\Omega_0u_3,\\
   &\\ \allowdisplaybreak
C_{-1}&=\Omega_2u_{-3}-4\Omega_1u_{-2}+6\Omega_0u_{-1}-4\Omega_{-1}u_0+
\Omega_{-2}u_1\\
C_0&=\Omega_2u_{-2}-4\Omega_1u_{-1}+6\Omega_0u_0-4\Omega_{-1}u_1+
\Omega_{-2}u_2\\
C_1&=\Omega_2u_1-4\Omega_1u_0+6\Omega_0u_1-4\Omega_{-1}u_2+\Omega_{-2}u_3.
\endalign
$$

The commutation relations in $\mho(\sLtwo,\pi_3)$ have the form
$$\allowdisplaybreaks\align
[u_{-3},u_{-2}]&=3A_{-5}\\
[u_{-3},u_{-1}]&=6A_{-4}\\
[u_{-3},u_0]&=9A_{-3}-\tfrac32D_{-3}\\
[u_{-3},u_1]&=12A_{-2}-6D_{-2}\\
[u_{-3},u_2]&=15A_{-1}-6D_{-1}-\tfrac37C_{-1}\\
[u_{-3},u_3]&=18A_0-42D_0-\tfrac{36}7C_0\\
[u_{-2},u_{-1}]&=3A_{-3}+D_{-3}\\
[u_{-2},u_0]&=6A_{-2}+\tfrac32D_{-2}\\
[u_{-2},u_1]&=9A_{-1}-3D_{-1}+\tfrac17C_{-1}\\
[u_{-2},u_2]&=12A_0-11D_0+\tfrac47C_0\\
[u_{-2},u_3]&=15A_1-6D_1-\tfrac37C_1\\
[u_{-1},u_0]&=3A_{-1}+\tfrac32D_{-1}-\tfrac3{35}C_{-1}\\
[u_{-1},u_1]&=6A_0+2D_0-\tfrac4{35}C_{-1}\\
[u_{-1},u_2]&=9A_1-3D_1+\tfrac17C_1\\
[u_{-1},u_3]&=12A_2-6D_2\\
[u_0,u_1]&=3A_1+\tfrac32D_1-\tfrac3{35}C_1\\
[u_0,u_2]&=6A_2+\tfrac32D_2\\
[u_0,u_3]&=9A_3-\tfrac32D_3\\
[u_1,u_2]&=3A_3+D_3\\
[u_1,u_3]&=6A_4\\
[u_2,u_3]&=3A_5,
\endalign
$$
where $D_k=B_k-2u_k+\varkappa(4Ku_k+66u_k-15B_k)$ ($K$ is $\sLtwo$--Casimir
element: $K=L_{-1}L_1-2L_0^2+L_1L_{-1}$; the constant $\varkappa$ is fixed by
the Jacobi identities and maybe found verifying them for the triple $u_{-2}$,
$u_{-1}$, $u_0$).

\

{\bf Hypothesis 2.} The statement of the Theorem remains true for $n>3$.

\head TOPIC THREE: LIE $\Fg$--BUNCHES AND RELATED HIDDEN SYMMETRIES\endhead

\definition{Definition 3}

A. Let $\Fg$ be a Lie algebra. A {\it Lie $\Fg$--bunch\/} is a $\Fg$--module
$W$ such that there exists a $\Fg$--equivariant mapping
$\Fg\otimes\bigwedge^2(W)\mapsto W$, which defines a structure of Lie algebra
in
$W$ when the first argument is fixed in an arbitrary way; we shall denote this
mapping by $[\cdot,\cdot]_L$, $L\in\Fg$.

B. The family (linear space) of operators is called {\it the isocommutator
algebra\/} (or {\it Lie isoalgebra\/} [7]) {\it of operators\/} if there exist
an operator $A$, which is called {\it an isotopic element\/} or shortly {\it
isotopy\/}, such that for all $X$ and $Y$ from the family the expression
$XAY-YAX$ also belongs to the family\footnote{\ It should be mentioned that an
isocommutator algebra of operators is a Lie algebra as abstract one. The
correspondence of an isocommutator algebra of operators to the abstract Lie
algebra is called {\it an isorepresentation} of the least.}. It should be
mentioned that all isotopic elements for a fixed family of operators form a
linear space\footnote{\ If $A$ is an isotopic elements for a family (linear
space) of operators and $L$ is an operator such that $\ad L$ preserves this
family then $[L,A]$ is also an isotopic element. Also if $A$ and $B$ are two
isotopic elements and $X$ is an arbitrary operator from the family then
$AXB-BXA$ is also an isotopic element (such {\sl "isotopic duality"} claims a
special and very serious attention).}.

Let $\Fg$ be a Lie algebra and $H$ be its module, if $L$ belongs to $\Fg$
then its image in $\End(H)$ will be denoted by $\pi(L)$. The family (linear
space) of operators in $H$ invariant with respect to the adjoint action of
$\Fg$ is called {\it the isocommutator algebra\/} (or {\it Lie isoalgebra\/})
{\it of hidden symmetries\/} if it is the isocommutator algebra of
operators for an arbitrary $\pi(L)$, $L\in\Fg$ as an isotopic element.

C. Let $\Fg$ be a Lie algebra and $W$ be a Lie $\Fg$--bunch. {\it An
isorepresentation\/} of $W$ is a $\Fg$--equivariant mapping $T$ from $W$ to
$\End(H)$, where $H$ is a certain $\Fg$--module, such that
$T([X,Y]_L)=T(X)\pi(L)T(Y)-T(Y)\pi(L)T(X)$, in other words $T$ corresponds an
isocommutator algebra of hidden symmetries to the Lie $\Fg$--bunch $W$.
\enddefinition

It should be specially mentioned that the considered case is linear, so the
Lie $\Fg$--bunch maybe straightforwardly restored from an isocommutator
algebra of operators in a similar way as the abstract Lie algebra is restored
from a commutator algebra of operators. Otherwords, there is no problem of
the setting hidden symmetries free in such situation.

Let's consider several examples, $\Fg=\sLtwo$.

\remark{\bf Example 1} {\sl There exists the unique structure of Lie
$\sLtwo$--bunch in $\pi_1$}.

Namely, if one fixes a basis $m_i$ ($i=-1,0,1$) in $\pi_1$ such that
$L_i(m_j)=(i-j)m_{i+j}$ then the isocommutators in the Lie $\sLtwo$--bunch
will have the form

\

\centerline{$\aligned
[m_{-1},m_0]_{L_0}&=m_{-1}\\
[m_1,m_{-1}]_{L_0}&=0\\
[m_1,m_0]_{L_0}&=m_1
\endaligned$ $\quad$
$\aligned
[m_{-1},m_0]_{L_{-1}}&=0\\
[m_1,m_{-1}]_{L_{-1}}&=2m_{-1}\\
[m_1,m_0]_{L_{-1}}&=2m_0
\endaligned$ $\quad$
$\aligned
[m_{-1},m_0]_{L_1}&=2m_0\\
[m_1,m_{-1}]_{L_1}&=-2m_1\\
[m_1,m_0]_{L_1}&=0
\endaligned$}
\endremark

\remark{\bf Example 2} Let $\pi_{1/2}$ be the 2--dimensional fundamental
representation of $\sLtwo$ then operators from $\End(\pi_{1/2})$ form
naturally an isocommutator algebra of hidden symmetries. The corresponding Lie
$\sLtwo$--bunch is realized in the direct sum $\pi_0\oplus\pi_1$ of the
trivial and adjoint representations of $\sLtwo$. If one fixes a basis $m_i$ in
$\pi_1$ as above and an element $c$ of $\pi_0$ (which is mapped to the identity
via an isorepresentation) then the isocommutators in the Lie $\sLtwo$--bunch
will have the form

\

\centerline{$\aligned
[m_{-1},m_0]_{L_0}&=0\\
[m_1,m_{-1}]_{L_0}&=-\tfrac12c\\
[m_1,m_0]_{L_0}&=0\\
[c,m_1]_{L_0}&=-m_1\\
[c,m_0]_{L_0}&=0\\
[c,m_{-1}]_{L_0}&=m_{-1}
\endaligned$ $\quad$
$\aligned
[m_{-1},m_0]_{L_{-1}}&=0\\
[m_1,m_{-1}]_{L_{-1}}&=0\\
[m_1,m_0]_{L_{-1}}&=\tfrac12c\\
[c,m_1]_{L_{-1}}&=-2m_0\\
[c,m_0]_{L_{-1}}&=-m_{-1}\\
[c,m_{-1}]_{L_{-1}}&=0
\endaligned$ $\quad$
$\aligned
[m_{-1},m_0]_{L_1}&=-\tfrac12c\\
[m_1,m_{-1}]_{L_1}&=0\\
[m_1,m_0]_{L_1}&=0\\
[c,m_1]_{L_1}&=0\\
[c,m_0]_{L_1}&=m_1\\
[c,m_{-1}]_{L_1}&=2m_0
\endaligned$}
\endremark

\remark{\bf Example 3} {\sl There are exactly four types of structure of a
Lie $\sLtwo$--bunch in $\pi_1\oplus\pi_2$}.

{\it Case 1: Enlarged example 1}.

\

\centerline{$\aligned
[m_{-1},m_0]_{L_0}&=a m_{-1}\\
[m_1,m_{-1}]_{L_0}&=0\\
[m_1,m_0]_{L_0}&=a m_1\\
[c,m_1]_{L_0}&=0\\
[c,m_0]_{L_0}&=\mu c\\
[c,m_{-1}]_{L_0}&=0
\endaligned$ $\quad$
$\aligned
[m_{-1},m_0]_{L_{-1}}&=0\\
[m_1,m_{-1}]_{L_{-1}}&=2a m_{-1}\\
[m_1,m_0]_{L_{-1}}&=2a m_0\\
[c,m_1]_{L_{-1}}&=-2\mu c\\
[c,m_0]_{L_{-1}}&=0\\
[c,m_{-1}]_{L_{-1}}&=0
\endaligned$ $\quad$
$\aligned
[m_{-1},m_0]_{L_1}&=2a m_0\\
[m_1,m_{-1}]_{L_1}&=-2a m_1\\
[m_1,m_0]_{L_1}&=0\\
[c,m_1]_{L_1}&=0\\
[c,m_0]_{L_1}&=0\\
[c,m_{-1}]_{L_1}&=-2\mu c.
\endaligned$}

\

{\it Case 2: Example 2}.

\

\centerline{$\aligned
[m_{-1},m_0]_{L_0}&=0\\
[m_1,m_{-1}]_{L_0}&=\lambda c\\
[m_1,m_0]_{L_0}&=0\\
[c,m_1]_{L_0}&=-d m_1\\
[c,m_0]_{L_0}&=0\\
[c,m_{-1}]_{L_0}&= d m_{-1}
\endaligned$ $\quad$
$\aligned
[m_{-1},m_0]_{L_{-1}}&=0\\
[m_1,m_{-1}]_{L_{-1}}&=0\\
[m_1,m_0]_{L_{-1}}&=-\lambda c\\
[c,m_1]_{L_{-1}}&=-2d m_0\\
[c,m_0]_{L_{-1}}&=-d m_{-1}\\
[c,m_{-1}]_{L_{-1}}&=0
\endaligned$ $\quad$
$\aligned
[m_{-1},m_0]_{L_1}&=\lambda c\\
[m_1,m_{-1}]_{L_1}&=0\\
[m_1,m_0]_{L_1}&=0\\
[c,m_1]_{L_1}&=0\\
[c,m_0]_{L_1}&=d m_1\\
[c,m_{-1}]_{L_1}&=2d m_0
\endaligned$}

\

{\it Case 3}.

\

\centerline{$\aligned
[m_{-1},m_0]_{L_0}&=a m_{-1}\\
[m_1,m_{-1}]_{L_0}&=\lambda c\\
[m_1,m_0]_{L_0}&=a m_1\\
[c,m_1]_{L_0}&=0\\
[c,m_0]_{L_0}&=2a c\\
[c,m_{-1}]_{L_0}&=0
\endaligned$ $\quad$
$\aligned
[m_{-1},m_0]_{L_{-1}}&=0\\
[m_1,m_{-1}]_{L_{-1}}&=2a m_{-1}\\
[m_1,m_0]_{L_{-1}}&=2a m_0+\lambda c\\
[c,m_1]_{L_{-1}}&=-4a c\\
[c,m_0]_{L_{-1}}&=0\\
[c,m_{-1}]_{L_{-1}}&=0
\endaligned$ $\quad$
$\aligned
[m_{-1},m_0]_{L_1}&=2a m_0+\lambda c\\
[m_1,m_{-1}]_{L_1}&=-2a m_1\\
[m_1,m_0]_{L_1}&=0\\
[c,m_1]_{L_1}&=0\\
[c,m_0]_{L_1}&=0\\
[c,m_{-1}]_{L_1}&=-4a c
\endaligned$}

\

{\it Case 4}.

\

\centerline{$\aligned
[m_{-1},m_0]_{L_0}&=a m_{-1}\\
[m_1,m_{-1}]_{L_0}&=0\\
[m_1,m_0]_{L_0}&=a m_1\\
[c,m_1]_{L_0}&=-d m_1\\
[c,m_0]_{L_0}&=0\\
[c,m_{-1}]_{L_0}&=d m_{-1}
\endaligned$ $\quad$
$\aligned
[m_{-1},m_0]_{L_{-1}}&=0\\
[m_1,m_{-1}]_{L_{-1}}&=2a m_{-1}\\
[m_1,m_0]_{L_{-1}}&=2a m_0\\
[c,m_1]_{L_{-1}}&=-2d m_0\\
[c,m_0]_{L_{-1}}&=-d m_{-1}\\
[c,m_{-1}]_{L_{-1}}&=0
\endaligned$ $\quad$
$\aligned
[m_{-1},m_0]_{L_1}&=2a m_0\\
[m_1,m_{-1}]_{L_1}&=-2a m_1\\
[m_1,m_0]_{L_1}&=0\\
[c,m_1]_{L_1}&=0\\
[c,m_0]_{L_1}&=d m_1\\
[c,m_{-1}]_{L_1}&=2d m_0.
\endaligned$}
\endremark

\remark{\bf Example 4} Let $\CA$ be an arbitrary associative algebra with an
involution $*$, $\Fg=\{A\in\CA: A^*=-A\}$ is a Lie algebra. Let
$W=\{B\in\CA: B^*=B\}$ then there exists a natural structure of the
isocommutator algebra of hidden symmetries
in $W$.

\

In particular, let's considered $\CA$ being $\End(\pi_1)$,
$\Fg=\sothree\subset\CA$ -- the set of skew--symmetric matrices $3\times 3$
($\sothree\simeq\sLtwo$), $W$ -- the set of symmetric matrices $3\times 3$ (it
should be marked that $\sLtwo$ acts in $W$ and $W$ is isomorphic to
$\pi_0\oplus\pi_2$ as $\sLtwo$--module). So $W$ is an isocommutator algebra of
hidden symmetries and therefore $\pi_0\oplus\pi_2$ possesses a structure of
Lie $\sLtwo$--bunch.
\endremark

This construction of Lie $\sLtwo$--bunch maybe straightforwardly generalized
on the direct sums $\bigoplus^k_{i=1} \pi_{2i}\simeq S^2(\pi_k)$.

\remark{Remark} $\pi_2$ does not admit any structure of a Lie $\sLtwo$--bunch.
\endremark

\remark{{\bf Example 5} (A geometric example of Lie $\Fg$--bunch)} Let $\CM$
be a smooth manifold. $C^{\infty}(\CM)$ is a Lie $\Vect(\CM)$--bunch.

\definition{Definition 4}

A. The algebra $\FA$ is called {\it Lie--admissible\/} [8] if its commutator
algebra $\FA^{(-)}$ is a Lie algebra. The main example of Lie--admissible
algebras is an arbitrary associative algebra supplied by the operation
$(X,Y)\mapsto XRY-YSX$ (for arbitrary fixed elements\footnote{\ which are
called {\it genotopic elements\/} [7].} $R$ and $S$).

B. Let $\Fg$ be a Lie algebra. {\it A Lie--admissible $\Fg$--bunch\/} is a
$\Fg$--module $W$ such that there is defined a $\Fg$--equivariant mapping
$\Fg\otimes T^2(W)\mapsto W$ which defines a structure of Lie--admissible
algebra in $W$, when the first argument is fixed in an arbitrary way, we shall
denote this mapping by $\left<\cdot,\cdot\right>_L$, $L\in\Fg$. It will be
called {\it a Lie--admissible $\Fg$--bunch of the first kind\/} if the
operation $X\circ_L Y=\frac12(\left<X,Y\right>_L-\left<Y,X\right>_L)$ does not
depend on $L$ and {\it a Lie--admissible $\Fg$--bunch of the second kind\/} if
the operation $[X,Y]_L=\left<X,Y\right>_L-\left<Y,X\right>_L$ does not depend
on $L$.

A Lie--admissible $\Fg$--bunch $W$ will be called {\it quadratic\/} ({\it
homogeneous\/} or {\it nonhomogeneous}, respectively) iff (1) there exists a
$\Fg$--submodule $W_0$ ({\it coordinate base\/}) in $W$ such that the mapping
from $S^{\cdot}(W_0)$ into $W$ defined by the Weyl symmetrizations is a
$\Fg$--equivariant isomorphism; (2) for a basis $w_m$ in $W_0$ the following
relations hold $[w_i,w_j]_L=S_{ij}^{kl}w_k\circ_L w_l$ or
$[w_i,w_j]_L=S_{ij}^{kl}w_k\circ_L w_l+ R_{ij}^k w_k$, respectively.

C. Let $\Fg$ be a Lie algebra. {\it An isorepresentation\/} of a
Lie--admissible $\Fg$--bunch $W$ of the first kind is a $\Fg$--equivariant
mapping from $W$ to $\End(H)$, where $H$ is a certain $\Fg$--module
($\pi:\Fg\mapsto\End(H)$), such that
$T([X,Y]_L)=T(X)\pi(L)T(Y)-T(Y)\pi(L)T(X)$, $T(X\circ
Y)=\frac12(T(X)T(Y)+T(Y)T(X))$. {\it An isorepresentation\/} of a
Lie--admissible $\Fg$--bunch $W$ of the second type is a $\Fg$--equivariant
mapping from $W$ to $\End(H)$, where $H$ is a certain $\Fg$--module
($\pi:\Fg\mapsto\End(H)$), such that $T([X,Y])=[T(X),T(Y)]$, $T(X\circ_L
Y)=\frac12(T(X)\pi(L)T(Y)+T(Y)\pi(L)T(X))$.

\enddefinition

\remark{{\bf Example 6} (cf. Example 4 above)} Let $\CA$ be an arbitrary
associative algebra with an involution $*$, $\Fg=\{A\in\CA: A^*=-A\}$. The
space $W=\{B\in\CA: B^*=B\}$ possesses a natural structure of Lie--admissible
$\Fg$--bunch of the first kind, whereas $\Fg$ itself possesses a natural
structure of Lie admissible $\Fg$--bunch of the second kind.
\endremark

\

So a classical Lie algebra $\Fg$ is supplied by a natural structure of
Lie--admissible $\Fg$--bunch.

In particular, $\pi_1$ admits a structure of Lie--admissible $\sLtwo$--bunch
of the second kind, whereas $\pi_0\oplus\pi_2$ admits a structure of
Lie--admissible $\Fg$--bunch of the first kind. Moreover, structures of
Lie--admissible $\Fg$--bunches of the first kind are defined in
$\bigoplus^k_{i=0}\pi_{2i}$, whereas structures of Lie--admissible
$\sLtwo$--bunches of the second kind are defined in
$\bigoplus^k_{i=0}\pi_{2i+1}$.
\endremark

\

{\bf Question 2:} Are there any quadratic Lie--admissible $\sLtwo$--bunches of
the
first and second types with $\pi_2$ as a coordinate base, which admit
isorepresentations realizing elements of the coordinate base by the tensor
operators $D^2_i$ ($-2\le i\le 2$) in the Verma modules $V_h$?

\remark{{\bf Example 7} (A geometric example of Lie--admissible $\Fg$--bunch)}
Let $\CM$ be a smooth manifold. The space of differential operators
$\DOP(\CM)$ on $\CM$ is a Lie--admissible $\Vect(\CM)$--bunch of the first
kind.
\endremark

\head CONCLUSIONS \endhead

Thus, some curiosities of {\sl the inverse problem of representation theory}
were considered on simple finite dimensional examples related to the Lie
algebra $\sLtwo$. Such consideration maybe regarded as toy one for analogous
infinite dimensional problems in {\sl the modern quantum field theory}
(conformal field theory, integrable models, field theory in non--trivial
backgrounds, etc.; see f.e. [9]). Some constructions inspired by these topics
and related to infinite dimensional hidden symmetries, which are produced by
vertex operator fields, will be discussed in the forthcoming article [10].

\Refs
\roster
\item"[1]" Connes A., {\it Introduction \`a la g\'eom\'etrie noncommutative}.
InterEditions, Paris, 1990;\newline
Manin Yu.I., {\it Topics in non--commutative geometry}. Princeton Univ. Press,
Princeton, NJ, 1991.
\item"[2]" Juriev D., Setting hidden symmetries free by the noncommutative
Veronese mapping. J.Math.Phys. 35 (1994) -- to appear; hep-th/9402130.
\item"[3]" Vershik A.M., Algebras with quadratic relations. Selecta Math.
Sovietica 11 (1992) 293-315 (originally published in {\it Spectral theory of
operators and infinite--dimensional analysis}, Inst. Math., Kiev, 1984, p.32-57
[in Russian]).
\item"[4]" Karasev M.V., Maslov V.P., Naza\v\i kinski\v\i\ V.E., Algebras with
general commutation relations and their applications. I, II. Current Probl.
Math., Modern Achievements. Moscow, VINITI, V.13, 1979 [in Russian].
\item"[5]" Manin Yu.I., Some remarks on Koszul algebras and quantum groups.
Ann. Inst. Fourier 37(4) (1987) 191-205.
\item"[6]" Shelepin L.A., Clebsch--Gordan coefficient calculus and its
physical applications. In {\it Group--theoretical methods in physics}. Proc.
P.N.Lebedev Physical Institute (Moscow), V.70, 1973, p.3-120 [in Russian].
\item"[7]" Santilli R.M., {\it Foundations of theoretical mechanics. Vol.II.
Birkhoffian generalization of Hamiltonian mechanics}. Springer--Verlag, 1982.
\item"[8]" L\^ohmus J., Paal E., Sorgsepp L., {\it Nonassociative algebras in
physics}. Hadronic Press, Tarpon Springs, Florida, 1994.
\item"[9]" Cremmer E., Gervais J.-L., Roussel J.-F., The quantum group
structure of 2D gravity and minimal models. II. The genus--zero chiral
bootstrap. Commun. Math. Phys. 161 (1994) 597-630;\newline
Gervais J.-L., Roussel J.-F., Solving the strongly coupled 2D gravity. II.
Fractional spin operators and topological three--point functions. Preprint
LPTENS 94/01; hep-th/9403028.
\item"[10]" Juriev D., On infinite dimensional hidden symmetries, in
preparation.
\endroster
\endRefs
\enddocument